\documentclass[
    ,final            % use final for the camera ready runs
  ]
  {aipproc}

\layoutstyle{6x9}

\usepackage{bm}
\usepackage{amstext}
\usepackage{amsfonts}

\begin{document}

\title{Chronological null complete spacetimes admit a global time}

\classification{4.20.Aq; 04.20.Cv; 04.20.Dw; 04.20.Gz}
%                \texttt{http://www.aip..org/pacs/index.html}>}
\keywords      {Time function, stable causality, null
incompleteness}

%4.20.q Classical general relativity (see also 02.40.k Geometry, differential geometry, and topology)\\
%04.20.Cv Fundamental problems and general formalism \\
%04.20.Dw Singularities and cosmic censorship \\
%04.20.Ex Initial value problem, existence and uniqueness of solutions \\
%04.20.Fy Canonical formalism, Lagrangians, and variational principles \\
%04.20.Gz Spacetime topology, causal structure, spinor structure

\author{E. Minguzzi}{
  address={Dipartimento di Matematica Applicata, Universit\`a degli Studi di Firenze,  Via
S. Marta 3,  I-50139 Firenze, Italy. E-mail:
ettore.minguzzi@unifi.it} }

\begin{abstract}
 The result ``chronological spacetimes without lightlike
lines are stably causal'' is announced and motivated. It implies
that chronological spacetimes which are null geodesically complete
and satisfy the null genericity and the null (averaged) energy
condition admit a  time function.
\end{abstract}

\maketitle

%%%%%%%%%%%%%%%%%%%%%%%%%%%%%%%%%%%%%%%%%%%%
%% MAINMATTER
%%%%%%%%%%%%%%%%%%%%%%%%%%%%%%%%%%%%%%%%%%%%

\section{Introduction}

A {\em time function}, $t: M \to \mathbb{R}$, is a continuous
function on spacetime $(M,g)$ (a time oriented Lorentzian manifold)
which increases on every future directed causal curve. If it exists,
it provides a total ordering of the spacetime events which respects
the notion of causal precedence.

In the work ``The existence of cosmic time functions''
\cite{hawking68}, Hawking pointed out the equivalence between the
existence of a time function and the property of stable causality.
Recall that a spacetime is stably causal if the light cones can be
strictly widened everywhere without introducing closed timelike
curves. The notion of stable causality is often regarded as the
minimal causality requirement which allows to remove altogether from
the spacetime any form of causality violation -- from closed causal
curves, to almost closed chains of  causal curves. The result by
Hawking then proves the equivalence of two very desirable features
for a spacetime. Subsequent work has shown that the time function,
whenever it exists, can be chosen smooth with a timelike gradient
\cite{bernal04} (see also \cite{seifert77}).

The problem of reducing the existence of a time function to more
direct, physically reasonable conditions was not addressed by
Hawking and in fact it has remained open so far. In the conclusion
of their 1979 review ``Global structure of spacetimes''
\cite{geroch79} Geroch and Horowitz identified this problem as one
of the most important open problems concerning the global aspects of
general relativity together with that of proving the cosmic
censorship conjecture. In fact, the proof of the existence of a time
function may also be regarded as a first step towards the goal of
proving the global hyperbolicity of  spacetime starting from
physically well motivated assumptions.

This work announces a result which, as I shall argue, solves Geroch
and Horowitz's problem. In order to state this result let me recall
that a lightlike line is a inextendible achronal causal curve, in
particular a lightlike line is a lightlike geodesic without
conjugate points \cite[Chap. 10, Prop. 48]{oneill83}.
 In the forthcoming work \cite{minguzzi07d} I shall
prove the theorem (C): ``chronological spacetimes without lightlike
lines are stably causal''.  Here I shall just sketch the basic ideas
underlying the proof, and comment on the physical consequences of
this result.  A first important observation is that the mentioned
theorem involves only conformal invariant properties  and hence it
is a theorem on the causal structure of the spacetime.

Now, recall that if a spacetime is null geodesically complete,
satisfies the null genericity condition and the null convergence
condition then any inextendible lightlike geodesic admits a pair of
conjugate points \cite[Prop. 4.4.5]{hawking73} \cite[Prop.
12.17]{beem96}, and hence any such spacetime does not admit
lightlike lines. Finally, thanks to (C), provided the spacetime is
chronological, one gets that it is also stably causal and hence
admits a time function.

Physically this result is very satisfactory in that the null
convergence condition can even be replaced by the weaker averaged
null convergence condition,
\cite{tipler78,tipler78b,chicone80,borde87} which is the weakest
among the energy conditions which are usually imposed on the stress
energy (Ricci) tensor.

Note that the real universe could indeed be null geodesically
complete while being timelike geodesically incomplete. It is easy to
check that this case is compatible with all the singularity
theorems. Even if global hyperbolicity holds one cannot conclude
using Penrose's singularity theorem \cite{hawking73} (1965)
 that the presence of a trapped surface would lead
to a null incomplete geodesic, in fact this result holds only if the
Cauchy hypersurfaces are non-compact (and, by the way, if one
assumes global hyperbolicity there is no need to argue for the
validity of stable causality by assuming null geodesic
completeness). Thus the assumption of null geodesic completeness is
compatible with the constraints given by the theory and by the
observation. The mentioned theorem can then be used to substantiate
the existence of a time function on the physical ground of this mild
non-singularity requirement.

However, the theorem can also be used in the ``negative'' way as an
aid to singularity theorems because it proves that under the same
assumption of, say, Hawking and Penrose (1970) singularity theorem
\cite{hawking73}, the spacetime admits a time function and hence a
foliation of partial Cauchy hypersurfaces. Thus one of the boundary
assumptions of Hawking and Penrose (1970) singularity theorem,
namely the existence of a compact partial Cauchy surface, is truly
only a compactness requirement.

Finally, (C) can be regarded as a singularity theorem in its own
right, in fact in the form ``non-stably causal spacetimes either are
non-chronological or admit  lightlike lines'' receives the following
physical interpretation ``if there is a form of causality violation
on spacetime then either it is the worst possible, namely violation
of chronology, or the spacetime is singular'' a result which
clarifies the influence of causality violations on singularities.

\section{Sketch of the proof}

In this section I  motivate the claim (C): chronological spacetimes
without lightlike lines are stably causal. Recall Hawking's result
 that a chronological spacetime without lightlike lines is strongly
causal (1966 Adams prize essay, see \cite{hawking70}).  The proof of
(C) is based on the preliminary result that the absence of lightlike
lines implies the transitivity of the causal relation $\bar{J}^{+}$.
Take two pairs $(x,y) \in \bar{J}^{+}$ and $(y,z) \in \bar{J}^{+}$
and two sequences of causal curves $\sigma_n$ of endpoints $(x_n,
y_n) \to (x,y)$, and $\gamma_n$ of endpoints $(y'_n,z_n) \to (y,z)$.
The limit curve theorem states that each sequence has a subsequence
that either converges to a connecting causal curve or converges to a
(past in the $\sigma_n$ case, future in the $\gamma_n$ case)
 inextendible causal curve passing through $y$.

%
%\begin{figure}[ht] \psfrag{x}{$x$} \psfrag{y}{$y$} \psfrag{z}{$z$}
%\psfrag{xn}{$x_n$} \psfrag{yn}{$y_n$} \psfrag{yn2}{$y'_n$}
%\psfrag{zn}{$z_n$} \psfrag{EQ}{$U\times V \subset I^{+}$}
%\psfrag{bx}{$\bar{x}$} \psfrag{bxn}{$\bar{x}_n$}
%\psfrag{bz}{$\bar{z}$} \psfrag{z}{$z$}  \psfrag{bzn}{$\bar{z}_n$}
% \psfrag{sy1}{$\sigma$}
%\psfrag{sy2}{$\gamma$} \psfrag{g1}{$\sigma_n$}
%\psfrag{g2}{$\gamma_n$} \psfrag{U}{$U$} \psfrag{V}{$V$}
%\begin{center}
% \includegraphics[width=4.5cm]{stratproof2}
%\end{center}
%\caption{The proof of the transitivity of $\bar{J}^{+}$ in the
%potentially dangerous case. } \label{stratproof2}
%\end{figure}

\begin{figure}[ht]
%\begin{center}
 \includegraphics[width=4.5cm]{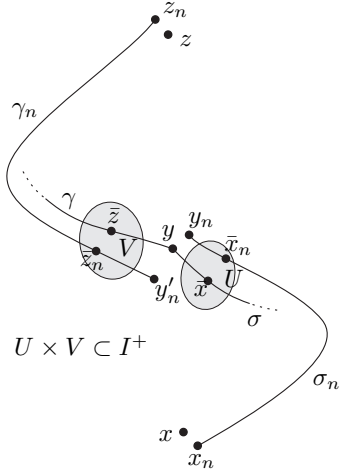}
%\end{center}
\caption{The potentially dangerous case in the proof of the
transitivity of $\bar{J}^{+}$. } \label{stratproof2}
\end{figure}

In the proof of $(x,z) \in \bar{J}^{+}$, the  only potentially
dangerous case is that in which neither subsequence converges to a
connecting curve (see figure \ref{stratproof2}). By joining at $y$
the two limit curves $\sigma$ and $\gamma$ one gets an inextendible
causal curve which by assumption is not a lightlike line. As a
result $\gamma \circ \sigma$ admits two chronologically related
events, $(\bar{x},\bar{z}) \in I^{+}$,  which without loss of
generality, can be found so that $\bar{x} \in \sigma$ and $\bar{z}
\in \gamma$. The fact that $I^{+}$ is open implies that a further
subsequence can be found such that $(x_{n_k},z_{n_k}) \in I^{+}$,
and hence $(x,z) \in \bar{J}^{+}$. With a similar argument it is
possible to show that the further assumption of chronology implies
not only that $(M,g)$ is strongly causal but also that the relation
$\bar{J}^{+}$ is antisymmetric (a property known as $A$-causality
\cite{woodhouse73,minguzzi07b}).

Now, note that if $\bar{J}^{+}$ is transitive then it is also the
smallest closed and transitive relation which contains $J^{+}$, that
is $K^{+}=\bar{J}^{+}$, where $K^{+}$ is the causal relation
introduced by Sorkin and Woolgar \cite{sorkin96}. A spacetime is by
definition, $K$-causal if $K^{+}$ is antisymmetric. Thus we have
shown that a chronological spacetime without lightlike lines is
$K$-causal.

It has long been suspected that $K$-causality may be equivalent to
stable causality. Indeed, R. Low \cite{sorkin96} suggested that
since stable causality is equivalent to the antisymmetry of the
Seifert relation \cite{seifert71} $J^{+}_S=\bigcap_{g'>g}
J^{+}_{g'}$ (a fact rigorously proved in \cite{hawking74} and
\cite{minguzzi07}), and this relation is closed and transitive, one
has $ K^{+}\subset J^{+}_S$, thus stable causality implies
$K$-causality, and maybe the equality $J^{+}_S=K^{+}$ holds  which
would imply that $K$-causality coincides with stable causality.
However, the situation proved more complex. Indeed, it was later
shown \cite{minguzzi07} that examples exist of spacetimes such that
$J^{+}_S\ne K^{+}$, but nevertheless no example is known of a
$K$-causal spacetime which is not stably causal.

%Indeed, evidence
% goes in the direction of confirming the
%equivalence and a theorem \cite{minguzzi07} states that if this
%coincidence holds then in any $K$-causal spacetime $K^{+}=J^{+}_S$.

The last step of the proof of (C) would be provided by the proof of
the equivalence between $K$-causality and stable causality. Recently
I gave a proof of this result \cite{minguzzi08b},  but for for the
sake of proving (C) it is possible to follow another simplified
route \cite{minguzzi07d} which avoids the direct proof of the
equivalence between stable and $K$-causality. In this strategy one
first define a property weaker that stable causality, which I termed
{\em compact stable casuality} and then takes advantage of the fact
that the property ``the spacetime is compactly stably causal and
does not have lightlike lines'' is invariant under suitable
enlargement of the light cones over compact sets \cite{minguzzi07d}.

% In
% I will follow another route, by avoiding such a
%direct proof. The reader is then referred to \cite{minguzzi07d} for
%the details, which involve a new causal property related but weaker
%than stable causality.

%%%%%%%%%%%%%%%%%%%%%%%%%%%%%%%%%%%%%%%%%%%%%%%%
%% BACKMATTER
%%%%%%%%%%%%%%%%%%%%%%%%%%%%%%%%%%%%%%%%%%%%%%%%

\begin{theacknowledgments}
  This work has been partially supported by GNFM of INDAM.
\end{theacknowledgments}



\bibliographystyle{aipproc}   % if natbib is available


%\bibliography{../../bibliografie/simultaneity,../../bibliografie/libri,../../bibliografie/miei,../../bibliografie/mieiPreprints,../../bibliografie/mieiProceedings}

\IfFileExists{\jobname.bbl}{}
 {\typeout{}
  \typeout{******************************************}
  \typeout{** Please run "bibtex \jobname" to obtain}
  \typeout{** the bibliography and then re-run LaTeX}
  \typeout{** twice to fix the references!}
  \typeout{******************************************}
  \typeout{}
 }

\end{document}